\documentclass[12pt]{iopart}
\usepackage{graphicx}

\begin{document}

\title[Geometric effects on critical behaviours of the Ising model]
{Geometric effects on critical behaviours of the Ising model}

\author{Hiroyuki Shima and  Yasunori Sakaniwa}

\address{Department of Applied Physics, Graduate School of Engineering,
Hokkaido University, Sapporo 060-8628, Japan}
\ead{shima@eng.hokudai.ac.jp}

\begin{abstract}
We investigate the critical behaviour of the two-dimensional Ising model defined 
on a curved surface with a constant negative curvature. 
Finite-size scaling analysis reveals that the critical exponents 
for the zero-field magnetic susceptibility and the correlation length 
deviate from those for the Ising lattice model on a flat plane.
Furthermore, when reducing the effects of boundary spins,
the values of the critical exponents tend to those derived from the mean field theory.
These findings evidence that the underlying geometric character 
is responsible for the critical properties the Ising model
when the lattice is embedded on negatively curved surfaces.
\end{abstract}

\submitto{\JPA}

\maketitle

\section{Introduction}

Scaling concept plays a vital role in describing critical phenomena
associated with a second-order phase transition.
The fundamental hypothesis states that,
in the vicinity of a critical point,
the largest length scale of the fluctuation of the order parameter diverges
and all length scales contribute with equal importance
\cite{Kadanoff, Wilson1}.
Theoretical arguments based on this hypothesis
explain why most thermodynamic quantities
near the critical point exhibit
power-law behaviour with characteristic exponents that are independent of
the microscopic details of a system
\cite{Cardy, Fisher_Review, Nakayama}.

A primary example of a physical model exhibiting a second-order phase transition
is the two-dimensional Ising model with ferromagnetic interaction
\cite{Onsager,Kaufmann,McCoy}.
This model has been used extensively for innumerable projects 
in statistical physics, mainly due to its simplicity and 
broad applicability to real systems.
Whereas its physical properties have been thoroughly investigated,
it still continues to raise interesting issues 
that are relevant to a wide range of critical phenomena.
Most intriguing among them is the critical properties of the Ising model 
defined on curved geometry
\cite{Izmailian_torus, sp_1, sp_2, sp_3, sp_4, sp_5, genus2}.
In fact, several studies have been carried out on Ising lattice models with topologies
ranging from a torus \cite{Onsager, Izmailian_torus, Ferdinand},
and sphere \cite{sp_1, sp_2, sp_3, sp_4, sp_5}
to genus two curved surfaces \cite{genus2},
as well as those with Brascamp-Kunz boundary conditions \cite{Janke, Izmailian_BK}.
The results of these studies
are consistent with the fact that an alteration in the topology or the boundary condition
of the Ising lattice
does not change its scaling behaviour,
and thus the system remains in the flat-space
Onsager universality class \cite{Onsager}.

The objective of the current study is 
to focus on an alternative property of curved geometries
--- surface {\it curvature} ---
instead of their {\it topology}.
Our main concern is whether a uniform change in the constant surface curvature
affects the scaling behaviour of the mounted Ising lattice model.
It should be noted that in most systems considered thus far,
the magnitude of surface curvature is
spatially concentrated on a portion of the surface,
even producing a conical singularity.
This inhomogeneous distribution of local curvature would possibly make it difficult
to distinguish the effect of curvature from among other incidental contributions
on the critical properties.
In addition, when a surface has a closed form (e.g., a sphere), 
its ability to attain the thermodynamic limit with a constant surface curvature
is disabled \footnote{For example, 
a sphere reduces to a flat plane at some point 
within the thermodynamic limit where the effect of curvature is absent.}.
This limitation can be removed successfully by employing a surface 
with a constant negative curvature.
This surface, in which the Gaussian curvature possesses a finite constant value
at arbitrary points, is simply connected and infinite
\cite{Coxeter, Firby}.
Hence, such a surface can serve as an example for considering 
the geometric effects on the critical properties of the mounted system.

In the present paper, we investigate the critical behaviour of the two-dimensional
Ising lattice model defined on a curved surface with a constant negative curvature.
Monte Carlo (MC) simulations and finite-size scaling analyses
are employed to compute the critical exponent $\gamma$ for the zero-field
magnetic susceptibility and $\mu$ for the correlation volume.
We demonstrate that the values of both $\gamma$ and $\mu$ deviate
from those for the planar Ising lattice model,
which indicates the relevance of the intrinsic geometry of the underlying surface
to the critical properties of the mounted Ising model.
Moreover, when reducing the boundary contributions,
the values of $\gamma$ and $\mu$ exhibit a tendency to shift to those derived from 
the mean-field approximation.
This non-trivial behavior of the critical exponents
is qualitatively consistent with the conclusion based on the series expansion analyses \cite{Rietman}
and that deduced from the quantum field theory \cite{JSTAT}.
We also calculate the fourth-order Binder's cumulant 
that provides a check of the results of finite size scaling.

\section{Scaling arguments for the Ising model}

Let us briefly review the framework of the scaling argument 
that successfully explains the critical properties of 
the two-dimensional planar Ising lattice model \cite{Cardy}.
The scaling hypothesis states that,
in the vicinity of the critical temperature $T_{\rm c}$,
the singular part of the free energy $f_{\rm s}$ of the Ising lattice per site
should be a homogeneous function:
\begin{equation}
\lambda^d f_{\rm s}(t,h) = f_{\rm s} (\lambda^x t, \lambda^y h), 
\label{eq_01}
\end{equation}
where $t=(T-T_{\rm c})/T_{\rm c}$, $h=H/(k_B T)$ and $H$ represents
an external magnetic field.
The parameter $\lambda^d$ with a spatial dimension $d$
indicates the rescaling of the total number of sites
from $N$ to $\lambda^d N$;
this results from the transformation of the linear dimension of 
the entire Ising lattice: $L\to \lambda L$.
By eliminating $\lambda$ from (\ref{eq_01}),
we obtain the scaling relation of $f_{\rm s}$:
\begin{equation}
f_{\rm s} \left(t, h \right) 
= \left| t \right|^{d/x} {\cal F} \left( \frac{h}{|t|^{y/x}} \right),
\label{eq_02}
\end{equation}
where ${\cal F}$ is a universal scaling function.
The appropriate differentiation of (\ref{eq_02}) yields
the power-law form of thermodynamic quantities
such as the zero-field susceptibility 
$\chi\equiv \partial^2 f_{\rm s}/\partial h^2 \propto |t|^{-\gamma}$,
where the critical exponents are expressed as functions of $x$ and $y$
(for instance, $\gamma = (2y-d)/x$).

Equation (\ref{eq_01}) is justified when the Ising lattice is defined
on a flat plane,
since the rescaling of $L$ by $\lambda$ is equivalent to 
that of $N$ by $\lambda^d$.
However, this is not the case when the Ising lattice is defined on a curved surface;
while a wide range of regular lattices can be constructed 
on curved surfaces with a constant Gaussian curvature
\cite{Coxeter, Firby},
the relation $N=L^d$ become invalid for these lattices due to the differences 
in the metric of the underlying geometry.
Hence, the rescaling $L\to \lambda L$ does not imply
$N\to \lambda^d N$,
which requires some modifications of (\ref{eq_01}).
\footnote{Similar argument has been made regarding
an infinitely coordinated Ising model; see Ref.~\cite{Botet}.}

We thus introduce an alternative rescaling parameter $\Lambda$
for considering the scaling relation of the Ising model on curved surfaces:
\begin{equation}
\Lambda f_{\rm s}(t,h) = f_{\rm s} (\Lambda^{\tilde{x}} t, \Lambda^{\tilde{y}} h),
\label{eq_03}
\end{equation}
where $\Lambda$ represents the rescaling of total sites $N\to \Lambda N$.
When the underlying geometry of the lattice is flat,
the relation (\ref{eq_03}) reduces to (\ref{eq_01})
since $\Lambda = \lambda^d$.
In this case, the parameters $\tilde{x}$ and $\tilde{y}$ given in (\ref{eq_03})
are defined as $\tilde{x}=x/d$ and $\tilde{y}=y/d$,
thus yielding identical values of critical exponents
(for instance, $\tilde{\gamma}=(2\tilde{y}-1)/\tilde{x}=(2y-d)/x = \gamma$).
In contrast, when the underlying geometry is curved,
$\Lambda$ can be no longer expressed as a power of $\lambda$;
thus, $\tilde{x}$ and $\tilde{y}$ are not related to $x$ and $y$.
Consequently, the critical exponents for the latter model,
which are determined by $\tilde{x}$ and $\tilde{y}$,
may quantitatively differ from those for the planar Ising lattice.
This naturally motivates us to evaluate the critical exponent directly
by constructing the Ising lattice model on curved surfaces.

\section{Regular tessellation of curved surfaces}

\begin{figure}[ttt]
\begin{center}
\includegraphics[width=5cm]{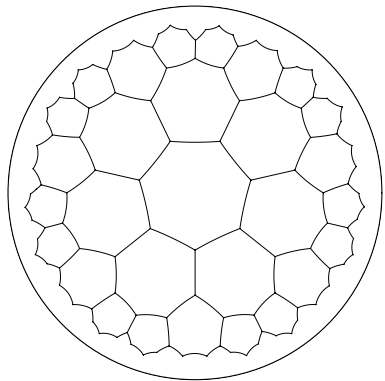}
\end{center}
\caption{Schematic illustration of 
a regular heptagonal lattice in terms of the Poincar\'e disk representation.
The number of concentric layers of heptagons is $r=3$ in this figure.
All heptagons depicted within the circle
are congruent with respect to the metric given in Eq.~(\ref{eqa_03}).
The circumference corresponds to an infinite distance from the center of the circle.}
\label{fig_1}
\end{figure}

A simple spherical surface seems to be the optimal geometry
to consider the curvature effect on the critical properties of the Ising model.
In fact, a number of efforts have been preformed 
on the Ising model with lattices whose topology is equivalent to a spherical surface
\cite{sp_1, sp_2, sp_3, sp_4, sp_5}.
It is noted that, however, 
the thermodynamic limit cannot be considered 
for the closed form of sphere-like surfaces having
{\it positive} curvature while maintaining their finite curvature.
This is because a spherical surface reduces to a flat plane in this limit, 
where the curvature effect vanishes completely.

Therefore, instead of a sphere,
we consider a curved surface with {\it negative} constant curvature, 
termed {\it a pseudosphere} \cite{Coxeter, Firby}.
The pseudosphere is a simply connected infinite surface
in which the Gaussian curvature at arbitrary points
possesses a constant negative value.
(The definition of the pseudosphere will be given in Appendix.)
Hence, it serves as a suitable geometry
for considering the curvature effect on the critical properties
of a system.
It should be noted that the pseudosphere occurs
in manifold physical problems ranging from quantum Hall effects
\cite{qhe1, qhe2, qhe3, qhe4, qhe5},
quantum chaos \cite{Balazs, qc2, qc3},
the string theory \cite{string}
to cosmology \cite{Levin},
wherein the underlying geometric character of the system
is extremely significant.

\begin{figure}[ttt]
\begin{center}
\includegraphics[width=7.5cm]{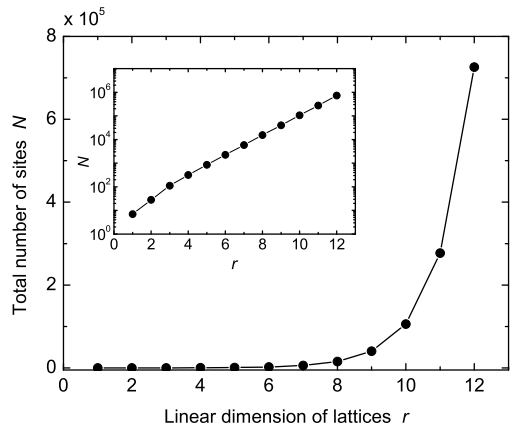}
\end{center}
\caption{Total number of sites $N$ 
involved in the heptagonal lattice with various system sizes.
The horizontal axis represents the number of concentric layers of heptagons,
$r$; this effectively serves as the linear dimension of the entire lattice.
Inset: A single logarithm plot of the identical data shown in Fig.~2.
It is clearly seen that $N$ for $r\gg 1$
increases exponentially with $r$.}
\label{fig_2}
\end{figure}

Interestingly, a wide range of regular lattices can be constructed on the pseudosphere
\cite{Firby}.
This is achieved by a tessellation procedure,
where the entire surface is covered 
by non-overlapping regular polygons meeting only along complete edges or at vertices.
It is known that a regular tessellation of the pseudosphere with $q$
regular $p$-sided polygons meeting at each vertex satisfies the following property:
\cite{Firby}
\begin{equation}
(p-2)(q-2) > 4.
\label{eq_04}
\end{equation}
Hence, the series of integer sets $\{p,q\}$ satisfying (\ref{eq_04})
results in an infinite number of possible regular tessellations of a pseudosphere.
This is in contrast to the case of a flat plane,
where only three regular tessellations are allowed:
$\{p,q\}=\{3,6\}$, $\{4,4\}$ and $\{6,3\}$
satisfying the condition $(p-2)(q-2)=4$.
For simplicity, 
we adopted a heptagonal $\{7,3\}$ tessellation to construct the Ising lattice
on a pseudosphere.
Figure 1 illustrates the local bond structure of 
a regular heptagonal lattice in terms of the Poincar\'e disk representation.
The resulting lattice comprises concentric layers of congruent heptagons
surrounding a central heptagon.
The Ising lattice models embedded on a pseudosphere have been considered thus far
\cite{Rietman, Elser, Angles};
however, explicit temperature dependences of thermodynamics quantities close to the transition
are yet to be concerned.

Due to a peculiar metric of a pseudosphere,
the total number of sites $N$ of our heptagonal lattice
exhibit a non-trivial evolution behaviour with the increase in the lattice size.
The size of our lattice is determined by 
the number of concentric layers of heptagons, denoted by $r$,
which effectively serves as a linear dimension in our lattice.
For a given $r$, the number of total sites $N$ is expressed as follows:
\begin{eqnarray}
N(r=1) &=& 7, \nonumber \\
N(r\ge 2) &=& 7+7 \sum_{j=0}^{r-2} 
\left[
c_+ \left( \frac{1+c_+}{2} \right)^{j}
+
c_- \left( \frac{1+c_-}{2} \right)^{j}
\right],
\label{eq_05a}
\end{eqnarray}
where $c_{\pm}=2 \pm \sqrt{5}$.
Figure \ref{fig_2} plots the dependence of $N$ on the effective linear dimension $r$.
When $r\gg 1$, it is approximated as
$N(r)\simeq 5 \exp(r)$; this means that
$N$ rapidly increases with $r$ in comparison with the case of the planar Ising model.
The exponential increase in $N(r)$
is a manifestation of the constant negative curvature of the underlying geometry
of our lattice.

It should be noted that, when considering thermodynamic properties of our lattice,
careful treatments on boundary effects are required.
The exponential increase in $N(r)$ results in that the ratio $[N(r)-N(r-1)]/N(r)$ approaches
a non-zero constant $1-e^{-1}$ in the limit $r\to \infty$.
This means that the boundaries of our lattices can not be neglected
even in the thermodynamic limit,
but contain a finite fraction of the total sites.
Boundary effects coming from these sites
are difficult to be eliminated completely,
because the periodic boundary conditions are hard to be employed to regular
lattices assigned on a pseudosphere.

In order to extract the bulk critical phenomena, therefore,
we have followed the procedure mentioned below.
Suppose that an Ising lattice consists of $r_{\rm out}$
concentric layers of heptagons.
Then, for computing physical quantities of the system
(magnetic susceptibilities, for instance),
we take into account only the Ising spins involved in
the interior $r_{\rm in}$ layers ($r_{\rm in} \le r_{\rm out}$)
so as to reduce the contribution of the spins locating near the boundary.
In actual calculations, $r_{\rm in}$ is varied from 4 to 8,
and for each $r_{\rm in}$ 
the number of disregarded layers 
$\Delta r \equiv r_{\rm out}- r_{\rm in}$ is systematically increased from 0 to 4.
By investigating the asymptotic behavior of the system for large $\Delta r$,
we can deduce the bulk properties of the Ising lattice model
embedded on the pseudosphere.

%
\section{Numerical methods}
%

We considered the conventional Ising model 
with ferromagnetic interaction:
\begin{equation}
H = -J \sum_{<i,j>} s_i s_j,\quad s_i=\pm 1,
\label{eq_05}
\end{equation}
where $\langle i,j \rangle$ denotes a pair of nearest-neighbour sites
on a heptagonal lattice.
The free boundary condition is imposed for all lattices to be considered.
Temperatures and energies are expressed as units of
$J/k_{\rm B}$ and $J$, respectively.
The order parameter $m$ per site for a given configuration of $\{s_i\}$
is given by $m=\sum_{i=1}^N s_i/N$.

Our main objective is to determine the zero-field magnetic susceptibility $\chi$
that exhibits the power-law relation $\chi(T)\propto |T-T_{\rm c}|^{-\gamma}$
near the critical temperature $T_{\rm c}$.
The susceptibility for a finite system size
can be expressed as a function of the order parameter $m$ as
\begin{equation}
\chi(T,N) = \frac{N \langle m^2 \rangle }{k_{\rm B}T},
\label{eq_06}
\end{equation}
or alternatively,
\begin{equation}
\chi'(T,N) = N\frac{\langle m^2 \rangle - {\langle |m| \rangle}^2}{k_{\rm B}T}.
\label{eq_07}
\end{equation}
Despite the difference in their definitions,
both $\chi$ and $\chi'$ yield the same critical exponent $\gamma$
by the finite-size scaling method, as described in Ref \cite{Binder}.
The expectation values $\langle |m| \rangle$
and $\langle m^2 \rangle$ at temperature $T$
are evaluated by canonical MC simulations \cite{Binder,Landau}.
Sampling of the configurational space was carried out
by using the single-cluster update algorithm \cite{Wolff}
which prevents the critical slowing down near the transition.

Quantitative evaluation of critical exponents
can be achieved by using the finite-size scaling technique
\cite{Finite, Cardy2, Privman}.
Close to the critical temperature $T_{\rm c}$,
the susceptibility for a finite system size satisfies
the following scaling behaviour:
\begin{equation}
\chi(T,N) \propto N^{\gamma/\mu}\cdot {\cal \chi}_0 \left( |T-T_{\rm c}|N^{1/\mu} \right).
\label{eq_07a}
\end{equation}
Here, $\mu$ is the critical exponent 
describing the divergence of the correlation volume 
$\xi_{\rm V}(T)$ of the order parameter:
%
%
\begin{equation}
\xi_{\rm V}(T)\propto |T-T_{\rm c}|^{-\mu}.
\label{eq_08}
\end{equation}
The quantity $\xi_{\rm V}$
is a natural generalization \cite{Botet, swn} of the correlation length $\xi$
that diverges as $\xi(T)\propto |T-T_{\rm c}|^{-\nu}$ with the critical exponent $\nu$
in the planar model.
Near $T_{\rm c}$,
the argument of the scaling functions ${\cal \chi}_0$ in (\ref{eq_07a}),
denoted by $x=|T-T_{\rm c}|N^{1/\mu}$,
becomes much smaller than unity.
This allows the polynomial expansion of the scaling function $\chi_0$ as
\begin{equation}
\chi(T,N) 
= 
a_0 N^{\gamma/\mu} + a_1 |T-T_{\rm c}|N^{(1+\gamma)/\mu} 
+ \cdots + a_n |T-T_{\rm c}|N^{(n+\gamma)/\mu},
\label{eq_10}
\end{equation}
terminating the expansion  at the order $n$.
By substituting the numerical data of $\chi(T,N)$ and their corresponding values of $T$ and $N$
into (\ref{eq_10}),
followed by performing the nonlinear least square fitting,
we obtain the critical exponents 
$\gamma$ and $\mu$ and the critical temperature $T_{\rm c}$ as optimal fitting parameters.

%
\section{Results}
%

\subsection{Susceptibilities and critical exponents for $\Delta r = 0$}

\begin{figure}[ttt]
\begin{center}
\includegraphics[width=6cm]{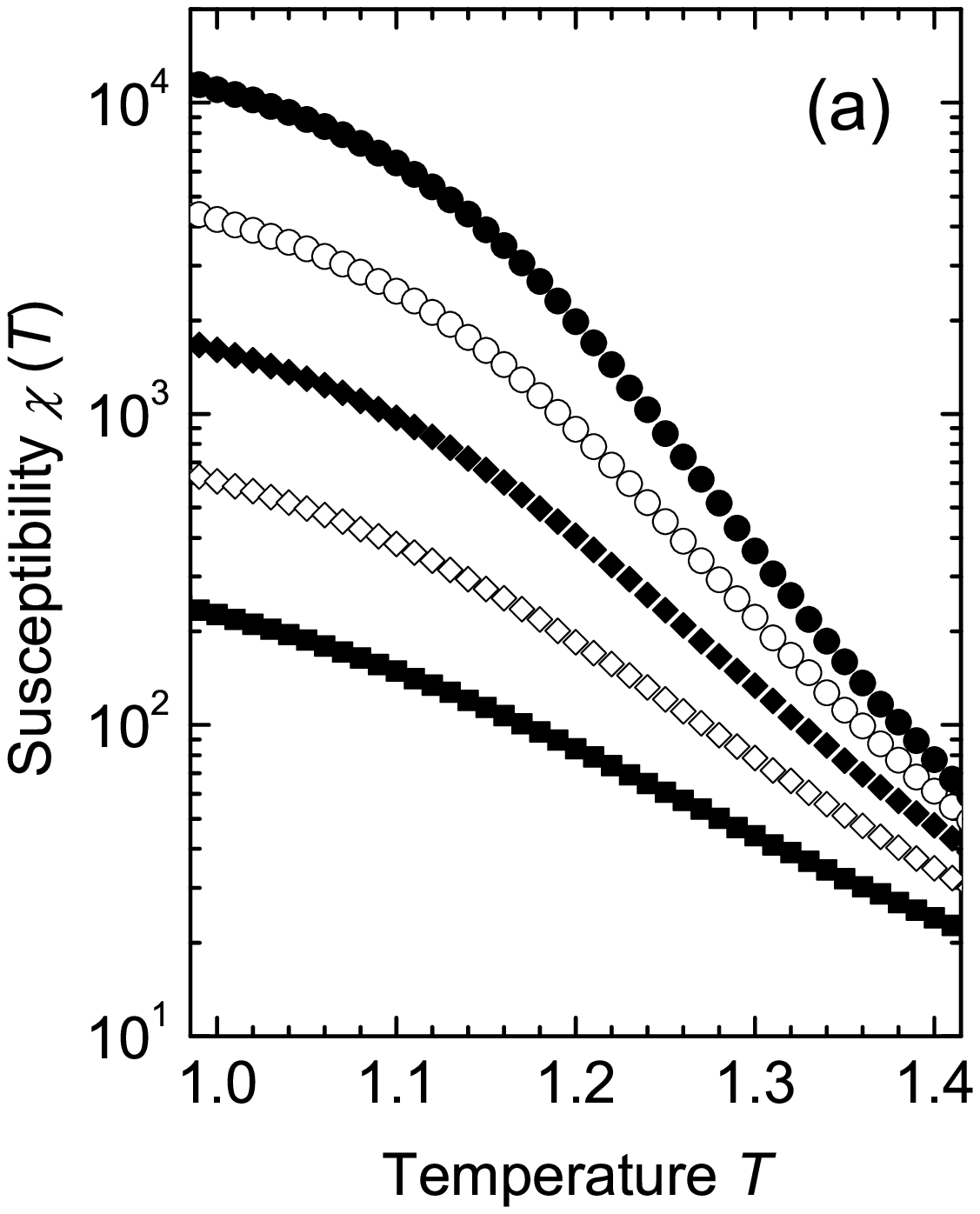}
\includegraphics[width=6cm]{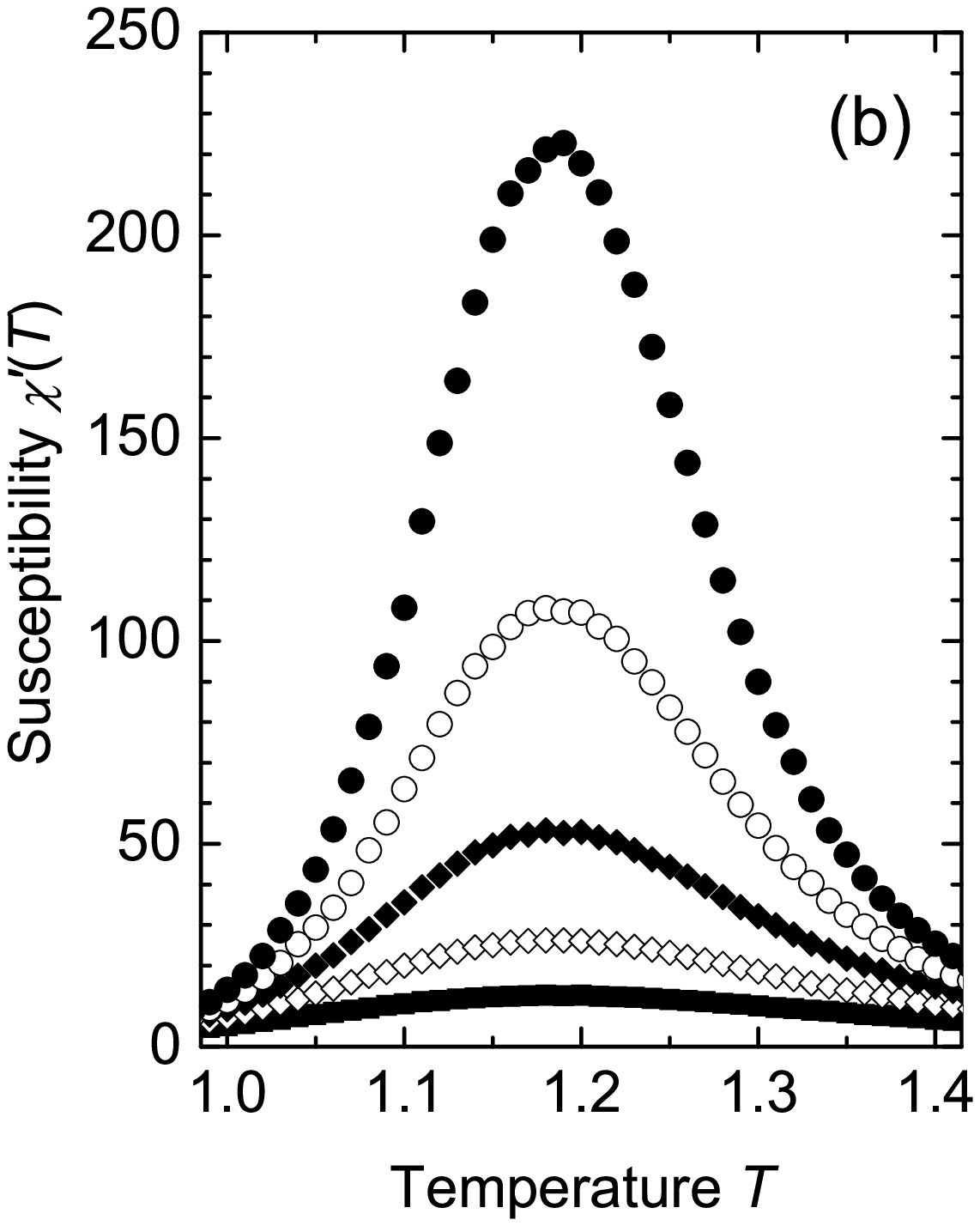}
\end{center}
\caption{(a) Zero-field magnetic susceptibilities:
(a) $\chi(T,N)$  and (b) $\chi'(T,N)$, for entire heptagonal lattices with $\Delta r = 0$.
The system size is varied from $r_{\rm in}=4$ (square) to $r_{\rm in}=8$ (solid circle),
which corresponds to the change in the number of total sites from $N=315$ to $N=15435$.}
\label{fig_3}
\end{figure}

Before addressing the bulk critical properties,
we first demonstrate the results for entire heptagonal lattices
with $\Delta r = 0$ (boundary contributions are fully involved).
Figure \ref{fig_3}(a) and \ref{fig_3}(b) show the calculated results of 
the zero-field susceptibilities $\chi(T,N)$ and $\chi'(T,N)$, respectively,
as a function of temperature $T$.
The single logarithmic plot is used in the figure \ref{fig_3}(a).
The system size $r = r_{\rm in}$ is varied from 4 to 8,
which corresponds to the change in the number of total sites from $N=315$ to $N=15435$.
Both $\chi$ and $\chi'$ exhibit such a typical behaviour
that indicates the occurrence of
a ferromagnetic transition within the temperature range of $1.1 \le T \le 1.3$.
For instance, the $\chi$ curve in figure \ref{fig_3}(a) monotonically increases 
with a decrease in $T$;
this is attributed to the onset of the ordered phase.
As well, the plot of $\chi'$ in figure \ref{fig_3}(b) exhibits a sharp peak at $T\sim 1.2$,
which is also a precursor of the divergence in the infinite system.

\begin{figure}[ttt]
\begin{center}
\includegraphics[width=6.0cm]{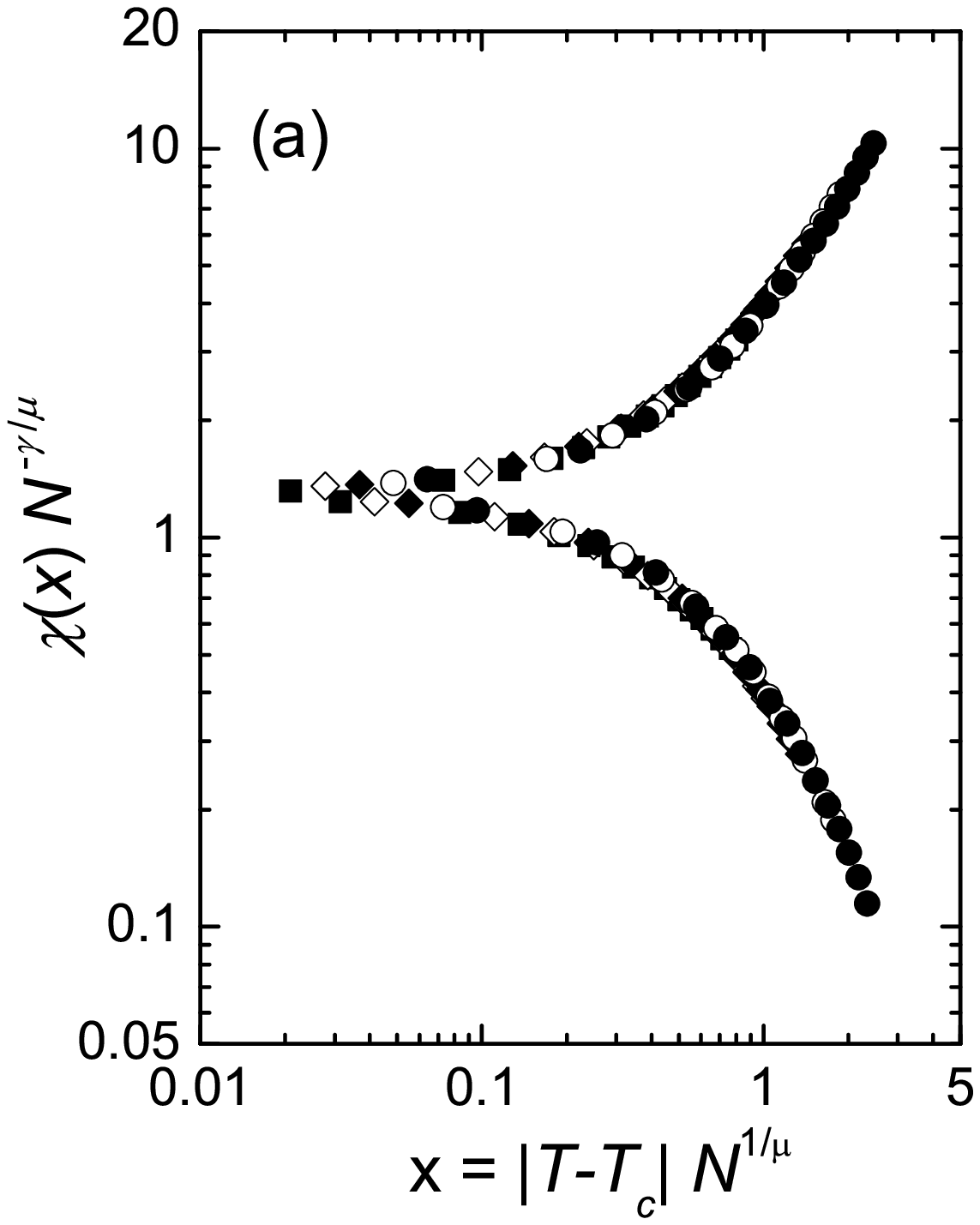}
\includegraphics[width=6.0cm]{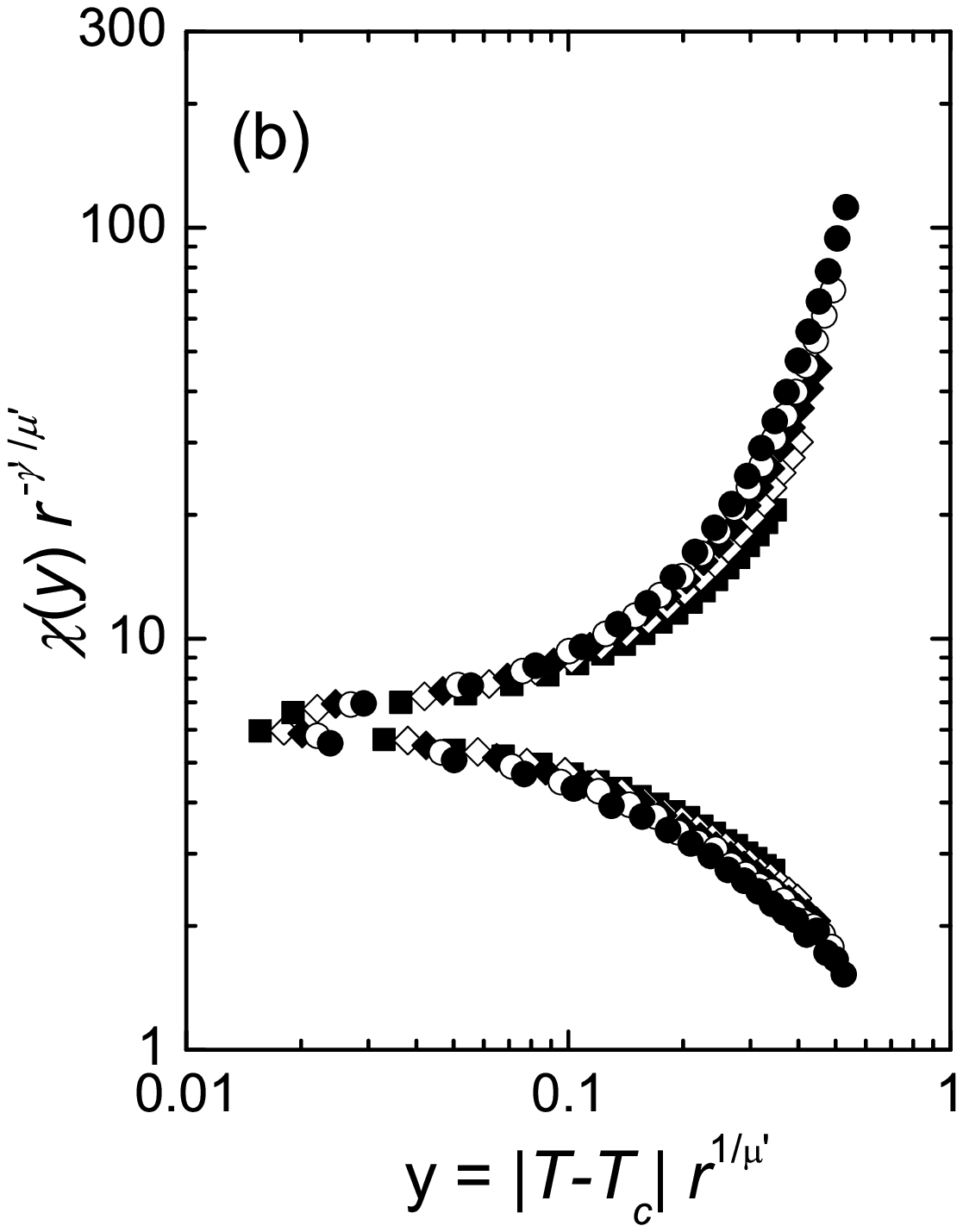}
\end{center}
\caption{(a) Scaling plot of $\chi(T,N)$
against the argument $x=|T-T_{\rm c}| N^{1/\mu}$.
The optimal values of critical exponents
are $\gamma=2.28(2)$ and $\mu=3.46(1)$,
and $\gamma=2.26(2)$ and $\mu=3.47(1)$
for the upper and lower branches, respectively.
The critical temperature 
is estimated as $T_{\rm c}=1.253(1)$ for the two branches.
(b) Scaling attempts of $\chi$ against the alternative argument 
$y=|T-T_{\rm c}| r^{1/\mu'}$ instead of $x$.
Optimal values of parameters are $\gamma'= 1.8(3)$, $\mu'=2.0(4)$, and $T_{\rm c}=1.46(4)$.}
\label{fig_4}
\end{figure}

Figure \ref{fig_4} (a) shows the scaling plot of $\chi(T,N)$ based on (\ref{eq_07a}).
The vertical and horizontal axes
represent the scaled susceptibility $\chi(x) N^{-\gamma/\mu}$
and its argument $x \equiv |T-T_{\rm c}| N^{1/\mu}$, respectively.
The critical exponents evaluated from the upper and lower branches are 
$\gamma=2.28(2)$ and $\mu=3.46(1)$,
and $\gamma=2.26(2)$ and $\mu=3.47(1)$, respectively.
The errors
in the last decimal places, which are shown in parentheses,
designate a 95\% confidence interval.
As expected, the estimated values of $\gamma$ and $\mu$
for the two branches are in agreement within numerical errors.
The optimal value of the critical temperature for the two branches
is evaluated as $T_{\rm c}=1.253(1)$;
this agrees with the preceding estimation from the values of 
figures \ref{fig_3}(a) and \ref{fig_3}(b).
A similar analysis for $\chi'$ results in 
$\gamma=2.26(3)$, $\mu=3.45(2)$, and $T_{\rm c}=1.254(2)$;
these values are fully consistent with the results deduced from the data of $\chi$.

We re-emphasize the fact that
for regular Ising lattices embedded on curved surfaces,
$N$ instead of $r$ should be adopted as the scaling variable.
This is justified by attempting the finite size scaling of $\chi$ using $r$.
Figure \ref{fig_4}(b) presents the scaling plot of $\chi$
in terms of another scaling argument $y=|T-T_{\rm c}| r^{1/\mu'}$;
the estimated values of the parameters are 
$\gamma'= 1.8(3)$, $\mu'=2.0(4)$, and $T_{\rm c}=1.46(4)$.
It is evident that the data points do not collapse onto a single curve,
but instead exhibit a large scatter.
Furthermore, the resulting value of $T_{\rm c}$ differs from the temperature
at which the susceptibility $\chi'(T)$ exhibits a sharp peak (see figure \ref{fig_3}).
These facts indicate that the linear dimension $r$ is not a characteristic length scale
that describes the scaling behaviour of the thermodynamic quantities for curved surfaces.

\subsection{Critical exponents for $\Delta r > 0$: Boundary effects}

We now turn to the study of bulk critical properties
of the heptagonal Ising lattice model.
As mentioned in Sec.~3, the boundary spins of Ising lattices on a pseudosphere
are thought to affect significantly to the nature of the system,
since the number of the spins along the boundary
increases as fast as that of total spins of the lattice.
Hence, in order to extract the bulk critical exponents,
we must try to remove the contribution of 
the boundary spins to the scaling behavior of the system.
This is achieved by setting the disregarded layers 
$\Delta r = r_{\rm out} - r_{\rm in}$ to be finite, i.e., 
by summing up only the spins within the interior $r_{\rm in}$ layers
when performing MC simulations
on the systems consisting of $r_{\rm out} (> r_{\rm in})$ layers.
If $\Delta r$ is sufficiently large,
the ensemble of the spins involved in the interior $r_{\rm in}$ layers
may yield the critical exponents
of the bulk system that is free from the boundary contribution.

\begin{figure}[ttt]
\begin{center}
\includegraphics[height=6.5cm]{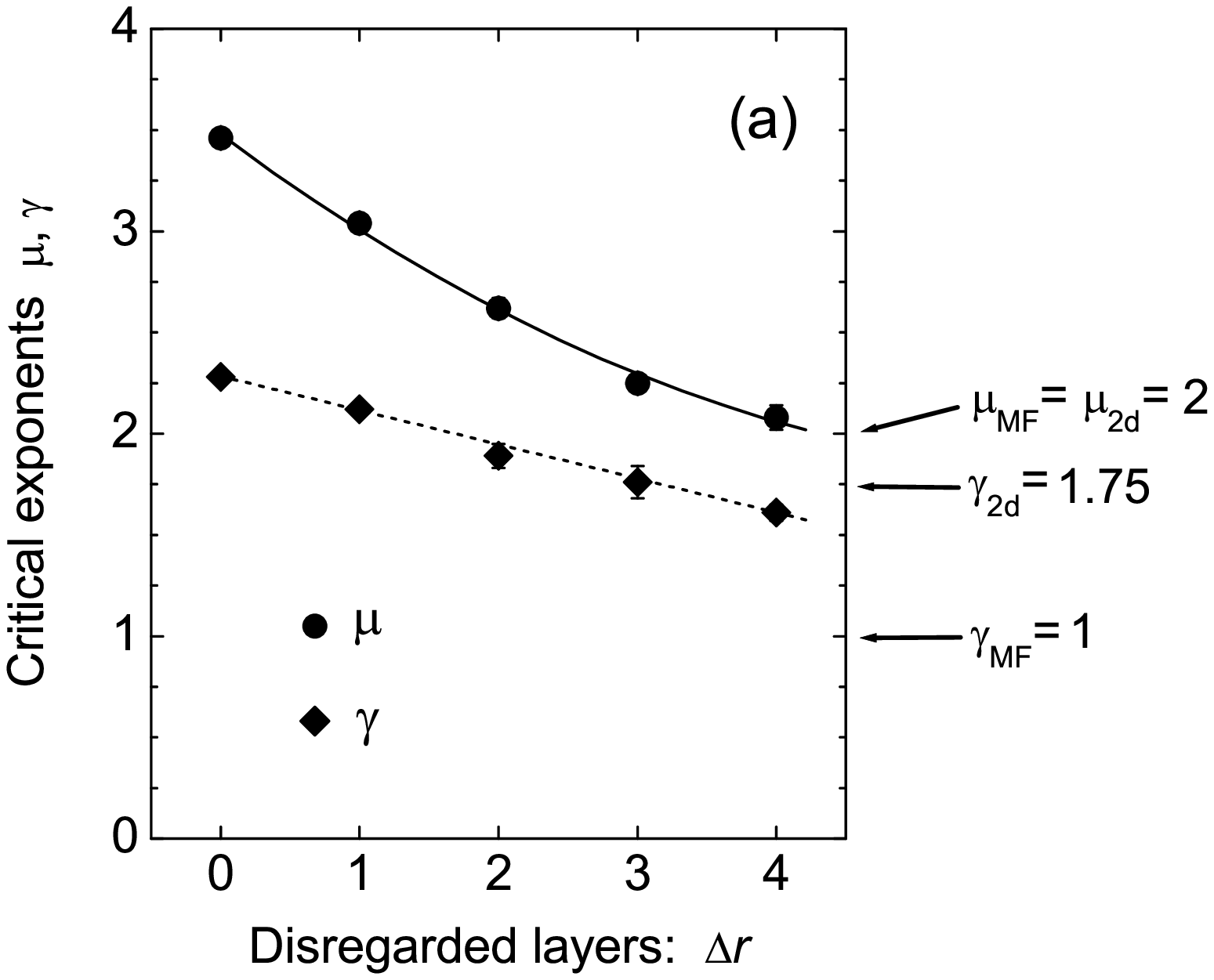}
\includegraphics[height=6.5cm]{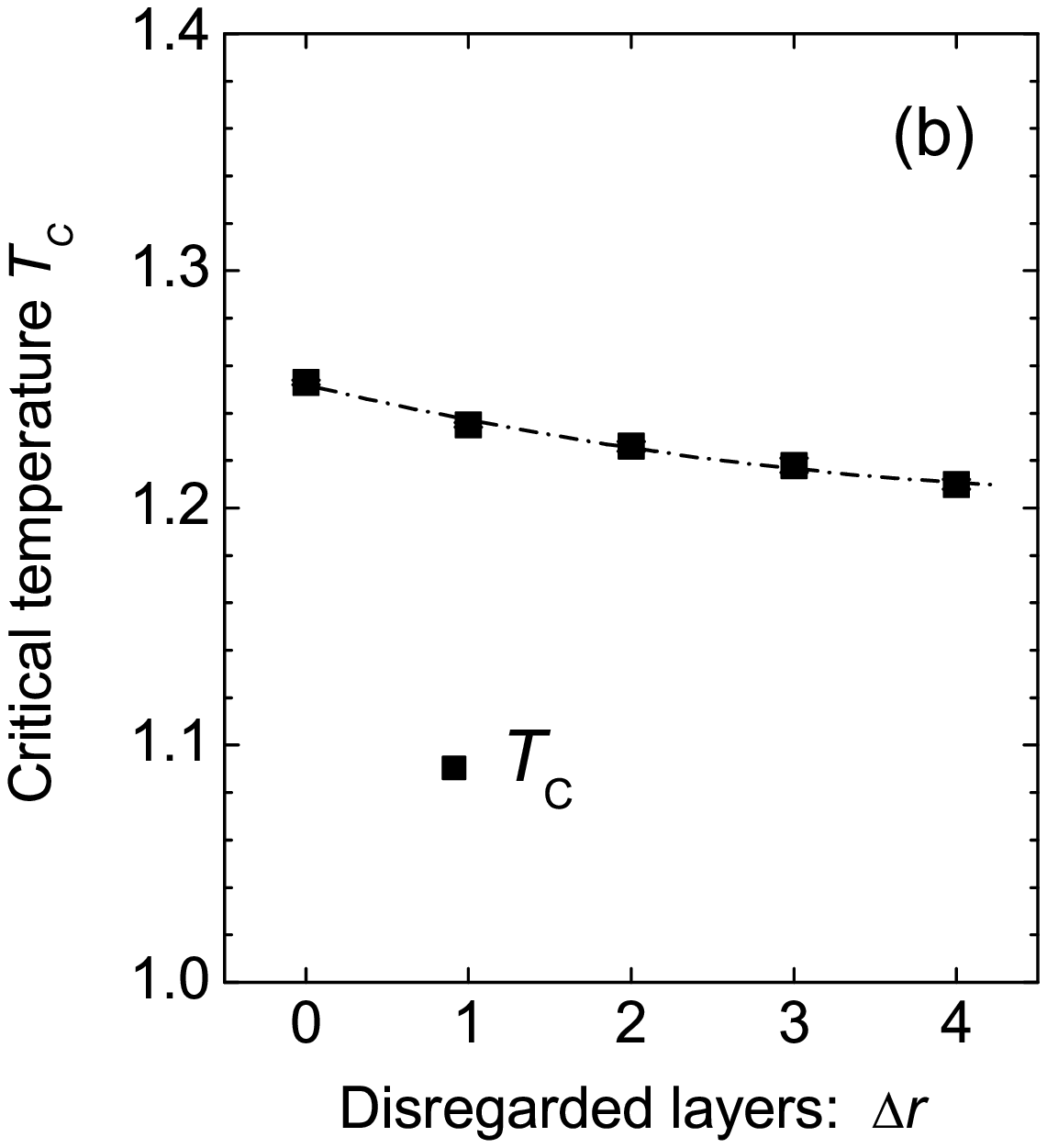}
\end{center}
\caption{(a) $\Delta r$-dependences of the critical exponents $\gamma$ and $\mu$,
and (b) that of the critical temperature $T_{\rm c}$.
The lines in the plots serve as a guide to eye.
Each data point was extracted by means of the finite size scaling analysis
for the system sizes $4\le r_{\rm in} \le 8$.
The values of the mean-field exponents $\gamma_{\rm MF}$ and $\mu_{\rm MF}$ are indicated
in the plot (a).}
\label{fig_5}
\end{figure}

On the basis of the argument above,
we have prepared the heptagonal lattices having various values of 
$r_{\rm in}$ and $\Delta r$,
and systematically employed the scaling analysis to them.
It then reveals how the critical exponents $\gamma$ and $\mu$
and the critical temperature $T_{\rm c}$ 
depend on the number of disregarded layers $\Delta r$.
The calculated results are given in Figure \ref{fig_5} (a) and \ref{fig_5} (b);
each data point in the plots was extracted by means of the finite size scaling analysis
for the system sizes $4\le r_{\rm in} \le 8$.
(Hence, the maximum system size we have treated reaches $r_{\rm out}=12$
which corresponds to $N=725760$.)
We found that an increase in $\Delta r$ results in a monotonic decrease
in all quantities in question: $\gamma$, $\mu$ and $T_{\rm c}$,
which indicates the significant contribution of the boundary spins
to the scaling behavior of the system.

Asymptotic behavior of the curves of $\gamma$ and $\mu$ for large $\Delta r$
provide estimations of the bulk critical exponents.
From Figure \ref{fig_5} (a), we see that the curve of $\mu$ shows somewhat convergence
to  a particular value of $\mu \sim 2$ or less.
Since this is close to the value of $\mu$ for the two-dimensional planar
Ising model\footnote{The value $\mu_{\rm 2d}=2$ is derived from the relation
$\mu=\nu d$ (See (\ref{eq_10})) and the exact solution $\nu=1$ for $d=2$.},
$\mu_{\rm 2d}=2$, 
it appears that the bulk system attains the planar Ising universality class
at around $\Delta r \sim 4$.
However, the asymptotic behavior of $\gamma$ is slightly different from that of $\mu$;
while the value of $\gamma$ equals to that of the planar system
$\gamma_{\rm 2d}=1.75$ at $\Delta r \sim 3$,
it still continues to decrease almost linearly with $\Delta r$
and thus has no tendency to converge to $\gamma_{\rm 2d}$ for large $\Delta r$.
Thereby, the bulk critical exponent $\gamma$ will be
smaller than $\gamma_{\rm 2d}$,
which suggests that the heptagonal Ising lattice
belongs to some other universality class
than of the planar Ising lattices.

In the context above, the asymptotic value $\mu\sim 2$ for large $\Delta r$
is not the exponent $\mu_{\rm 2d}$ but another specific exponent
characterizing the intrinsic nature of the system.
This point is clarified by referring
to the previous studies done by Rietman {\it et al.} \cite{Rietman}
and by Doyon and Fonseca \cite{JSTAT}.
They both have stated that the Ising lattice model embedded on the pseudosphere should yield
the mean-field critical exponents when the boundary contribution may be omitted.
The mean-field nature of the system is attributed to the fact that
an Ising lattice embedded on a pseudosphere is effectively an infinite dimensional lattice
at large distance due to the exponential growth of the total spins \cite{Callan}.
For an ordinary Ising lattice in $d$ dimension,
the number of spins along the boundary, $N_s$, 
is related to that of the total spins $N$ as $N_s \propto N^{1-(1/d)}$.
Hence, the peculiar relation $N_s \propto N$ that holds on negatively curved surfaces
consequences $d= \infty$.
Accordingly, our heptagonal Ising lattices are expected 
to yield the mean-field critical exponents $\gamma_{\rm MF} = 1$ and $\mu_{\rm MF}=2$,
where $\mu_{\rm MF}$ is determined by\footnote{For the Ising model,
the mean-field exponent $\nu_{\rm MF}$ is $1/2$, 
and the upper critical dimension $d_c$ is 4; thus we obtain $\mu_{\rm MF}=2$.}
$\mu_{\rm MF} = \nu_{\rm MF} d_c$ as suggested in Ref.~\cite{Botet}.
Our numerical results given in Figure \ref{fig_5} (a) is in fact 
qualitatively consistent with the argument above;
that is, $\mu$ converges to $\mu_{\rm MF}=2$ for $\Delta r \ge 4$,
and $\gamma$ continues to decrease until it yields $\gamma_{\rm MF}=1$.
(To be precise, the asymptotic value of $\gamma$ 
may take the value between $\gamma_{\rm 2d}$ and $\gamma_{\rm MF}$
depending on the numerical conditions;
this point will discussed in Section 6 in detail.)

\subsection{Binder's cumulant $U_4(T,N)$}

We have also carried out an alternative estimation of $T_{\rm c}$ and $\mu$
in terms of the fourth-order Binder's cumulant $U_4(T,N)$ defined by \cite{Binder, Landau, cumulant}
\begin{equation}
U_4(T,N) = 1 - \frac{\langle m^4\rangle}{3\langle m^2\rangle^2}.
\end{equation}
For a given $N$,
the cumulant $U_4(T)$ decreases monotonically with an increase in $T$
from $U_4(0)=2/3$ to $U_4(\infty)=0$. 
In the vicinity of $T_{\rm c}$, the $T$-dependence of it
can be approximated by
\begin{equation}
U_4(T,N) = U^{(0)} +  U^{(1)} \left( 1- \frac{T}{T_{\rm c}} \right) N^{1/\mu},
\label{eq_12}
\end{equation}
where $U^{(0)}$ and $U^{(1)}$ are constants and are thus
independent of $T$ and $N$.
The expression (\ref{eq_12}) is a direct consequence of the assumption
that the probability distribution of the order parameter $m$ should be Gaussian
close to the transition \cite{Binder}.
From (\ref{eq_12}), it follows
\begin{equation}
U_4(T_{\rm c}, N) = \mbox{const.}
\quad \mbox{and} \quad
\left. \frac{d U_4}{d T} \right|_{T=T_{\rm c}} \propto - N^{1/\mu},
\label{eq_14}
\end{equation}
which provides a complementary method to estimate $T_{\rm c}$ and $\mu$.

\begin{figure}[ttt]
\begin{center}
\hspace*{-1cm}
\includegraphics[height=6.5cm]{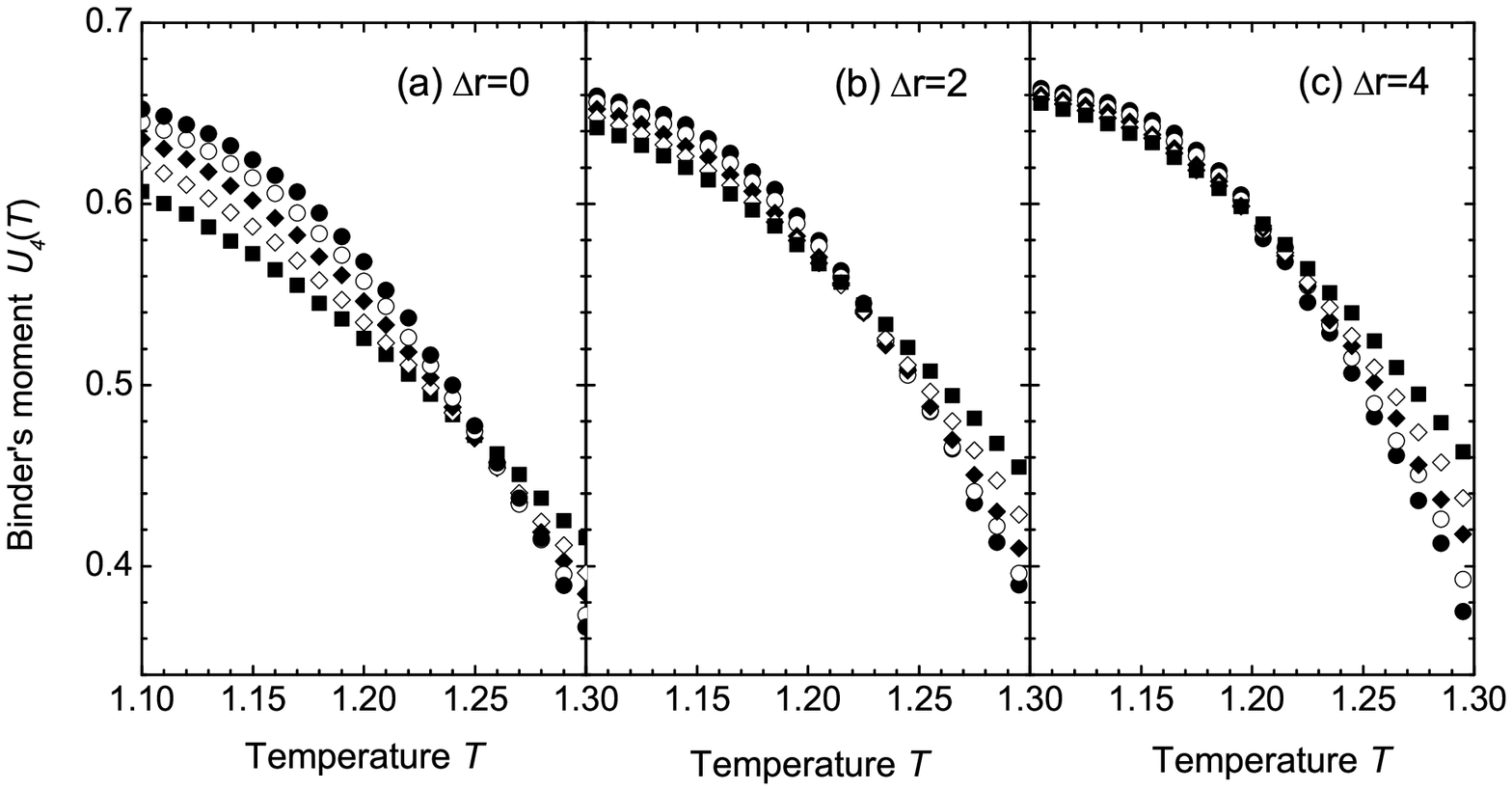}
\hspace*{-1.0cm}
\includegraphics[height=6.3cm]{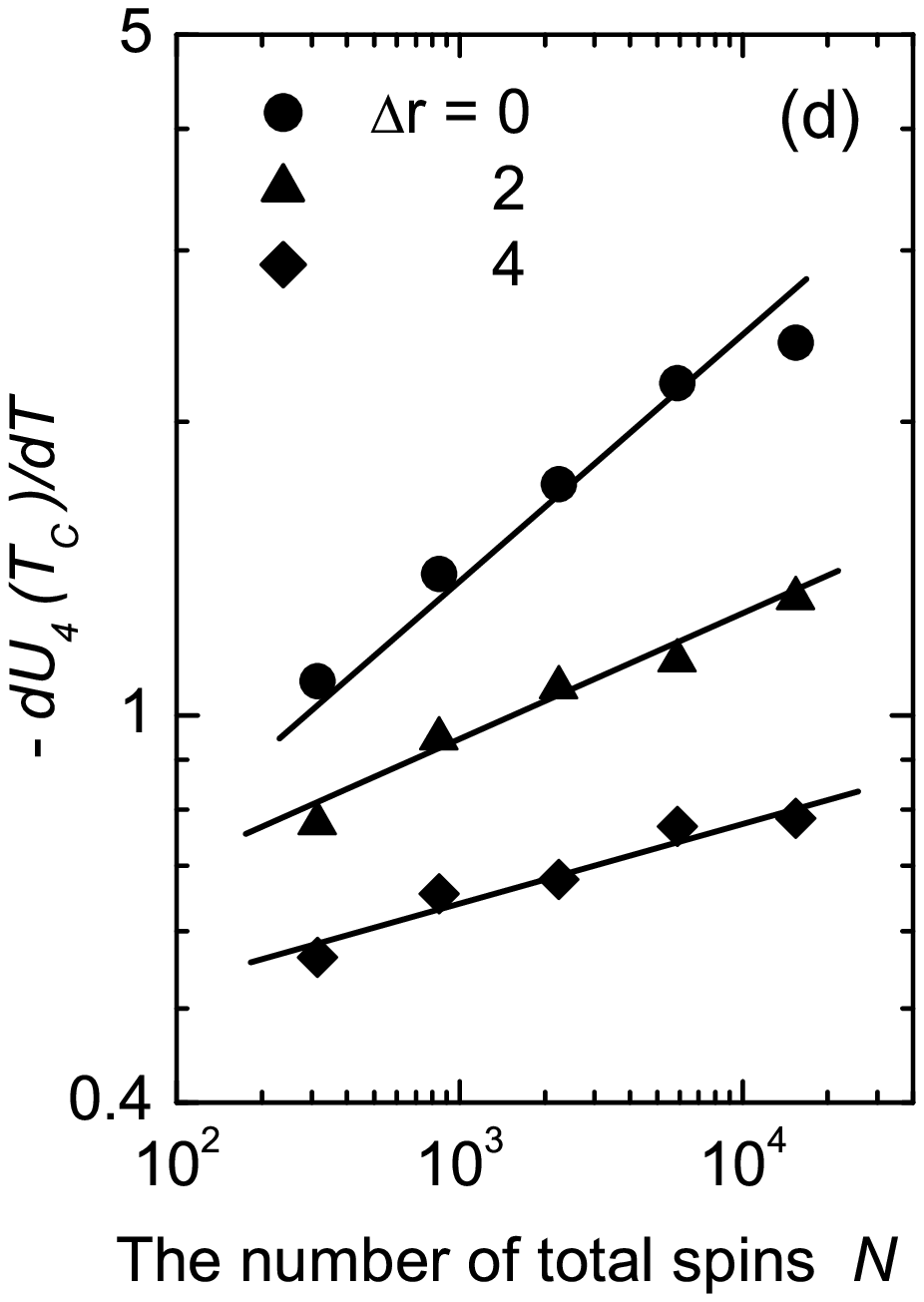}
\end{center}
\caption{(a)-(c) Temperature dependences of the 
fourth-order Binder's cumulant $U_4(T,N)$ for various values of $\Delta r$.
The size of the interior layers is varied from 
$r_{\rm in}=4$ (square) to $r_{\rm in}=8$ (solid circle).
(d) The negative slope of the cumulant, $-d U_4/d T|_{T=T_{\rm c}}$, 
for each value of $\Delta r$.
The solid lines represent the power-law $(\propto N^{1/\mu})$
with $\mu=3.6$, $7.7$, and $12.5$ from top to bottom.}
\label{fig_6}
\end{figure}

Figures \ref{fig_6} (a)-(c) present the numerical results for $U_4(T,N)$ for several
values of $\Delta r$.
In each plot, we found a unique crossing point giving an estimate of
the critical temperature as
$T_{\rm c} \simeq 1.25$, $1.22$ and $1.20$ for $\Delta r = 0$, $2$ and $4$, respectively.
These values of $T_{\rm c}$ are in fair agreement with those obtained by the scaling analyses
for the susceptibilities presented in Figure \ref{fig_5} (b).

Figure \ref{fig_6} (d) shows the $N$ dependence of
the derivative with negative sign $-d U_4/d T$ at $T=T_{\rm c}$
for different $\Delta r$.
The magnitude of the derivative grows with a power-law with increasing $N$ as expected
from (\ref{eq_14}).
The estimates of $\mu$ are $\mu \sim 3.6$, 7.7, 12.5
for $\Delta r=0$, 2, 4, respectively.
It should be noted that these values of $\mu$ increase with $\Delta r$.
This clearly contradicts to the results of the scaling analyses on the susceptibility $\chi$
(See Fig.~\ref{fig_5} (a)),
where $\mu$ is found to be such a decreasing function of $\Delta r$ that yields the mean-field
value $\mu_{\rm MF}=2$ in the large $\Delta r$ limit.
The discrepancy may be due to simply a finite sized effect;
or, it may indicate some intrinsic property of negatively curved surfaces
with regard to the distribution of the order parameter $m$,
since the power-law relation (\ref{eq_14}) originate from the assumption
of the Gaussian distribution of $m$ \cite{Binder}.
More through discussion about this point will be presented elsewhere.

%
\section{Discussions and Concluding remarks}
%

Our numerical analysis revealed that the critical exponents $\gamma$ and $\mu$
of the heptagonal Ising model defined on negatively curved surfaces
assume values that deviate from those for the planar Ising model.
Most striking is that, when reducing the contribution
of spins near the boundary,
both $\gamma$ and $\mu$ exhibit a tendency to yield the mean-field exponents.
This phenomenon is attributed to the fact that the regular Ising lattices
on negatively curved surfaces serve as an effectively infinite dimensional lattices
due to the peculiarity of the intrinsic geometry.

The above statement immediately poses the following question:
Does the negative curvature of the underlying geometry
alter the other four critical exponents?
With regard to the power-law behaviour of thermodynamic quantities,
the planar Ising model is known to possess
four other critical exponents \cite{Fisher_Review}:
$\alpha=0$, $\beta=1/8$, $\delta=15$, and $\eta=1/4$,
which correspond to heat capacity, spontaneous magnetization, critical isotherm,
and the two-point correlation function, respectively.
Our preliminary study \cite{ours} has suggested that,
the exponent $\beta$ for large $\Delta r$ also tend
toward the mean-field exponent $\beta_{\rm MF}=1/2$,
whereas $T_{\rm c}$ estimated there is slightly different from 
that in the present work.
Detailed analyses on this issue 
and the quantitative determination of the other exponent $\delta$ (and $\eta$,
if it exists\footnote{Rietman {\it et al.} \cite{Rietman}
have suggested that
the two-point correlation function of the Ising lattices on a pseudosphere
shows an exponential decay instead of an power-law decay.
If it is true, the critical exponent $\eta$ can be no longer defined.})
will be given in a future study.
Very recently, we have found that the dynamic critical exponent $z$
in our heptagonal Ising model also shift quantitatively
from that for the planar Ising model
\cite{dynamic}.

It deserves comment that the finite curvature of the underlying geometry 
may produce another type of effect wherein the spin variables 
at each site possess orientational degrees of freedom.
This is because
the relative angle of interacting spins at neighbouring sites
on a curved surface
is determined by a spatially-dependent metric tensor.
Thus, the Hamiltonian of the system should be modified
such that it is a function of the metric tensor.
As a result, the energetically preferable configurations of the vector spins 
differ from those in planar systems \cite{Hei1, Hei2, Hei3, Hei4, Hei5},
which implies that the critical behaviour of these vector-spin lattice models
is markedly influenced by a finite surface curvature.

Further noteworthy is, however, the fact that the geometric curvature 
continues to be relevant to the critical behavior of the system 
despite the omission of the vector property of the interacting spin variables; 
this was demonstrated by our results.
Evidently, the spin variable of our model is set to be a scalar,
and thus the Ising Hamiltonian given by (\ref{eq_05}) is devoid of the metric factor.
Nevertheless, the surface curvature is surely relevant to the Ising model on curved surfaces
since it enables to construct peculiar lattice structures
that cannot be realize in a flat plane.
This means that the surface curvature induces
the alteration of the global symmetry of the system
even when the interacting entities do not exhibit any vector property.
Besides, in the vicinity of the critical point, 
the discreteness of the lattice becomes irrelevant 
and the model can be considered to be a continuum surface with a constant curvature. 
Therefore, the metric of the underlying geometry plays a crucial role in the scaling behaviour
of the Ising lattice model defined on the surface.
In this context, the mean-field nature of the critical exponents $\gamma$ and $\mu$ 
are expected to be universal 
for all lattice structures other than the heptagonal one; 
this point is being investigated.

We remark that it is also interesting to study the dependence of the values of the bulk
critical exponents on the interior lattice size $r_{\rm in}$ we have introduced.
Obviously, the condition $1\ll r_{\rm in} \ll r_{\rm out}$ is desirable
to determine the bulk properties of the lattice with accuracy.
If the $r_{\rm in}$ is not so large (compared to the curvature radius of 
the underlying surface),
the system can be regarded as an Ising lattice defined on a nearly flat surface.
Thereby, the resultant critical exponents will become comparable with those for the planar
Ising lattice rather than the true bulk critical exponents (i.e., the mean-field exponents).
This implies that, by gradually increasing the size of $r_{\rm in}$,
the system go through a crossover from the planar Ising class
to the mean-field class.
We conjecture that this crossover phenomenon
is observed in the shift of the asymptotic value of $\gamma$
deduced from the plot in Figure \ref{fig_5} (a); that is,
when increasing the size of $r$ employed in the analyses,
the asymptotic value of $\gamma$ at $\Delta r \gg 1$
will shift downward from $\gamma\sim \gamma_{\rm 2d}=1.75$ to $\gamma \sim \gamma_{\rm MF}=1$.
Accordingly, it is possible that the curve of $\gamma$ plotted in 
Figure \ref{fig_5} (a) converges to an intermediate value 
between $\gamma_{\rm 2d}$ and $\gamma_{\rm MF}$ under the present numerical conditions.

In conclusion,
we have investigated the critical behaviour of the Ising model 
defined on a curved surface with constant negative curvature.
MC simulations and finite-size scaling analyses
were employed to compute the critical exponent $\gamma$ and $\mu$
for the zero-field magnetic susceptibility and correlation volume, respectively.
The resulting values $\gamma$ and $\mu$ show 
distinct values from those for the planar Ising model,
and exhibit a tendency to the mean-field exponents $\gamma_{\rm MF}=1$
and $\mu_{\rm MF}=2$ due to the peculiar intrinsic geometry of 
the negatively curved surface.
As well, we have revealed quantitatively how the boundary spins contribute
to the determination of $\gamma$, $\mu$ and $T_{\rm c}$,
and argued the possibility to occur the crossover from the planar Ising class
to the mean-field Ising class.
We hope that the generalization our statistical model
(with regard to the lattice structure, dimensionality of the embedding space,
distribution of interacting strength, {\it etc.})
would unveil a wide variety of interacting critical properties of the physical systems
assigned on general curved spaces.

\ack

We thank T.~Nakayama and K.~Yakubo for useful discussions.
This work was supported in part by a Grant--in--Aid for Scientific
Research from the Japan Ministry of Education, Science, Sports and Culture.
One of the authors (HS) is thankful for the financial support from 
the 21st Century COE ``Topological Sciences and Technologies,"
and for the conversation with S.~Tanda, Y.~Asano, and K.~Konno.
Numerical calculations were performed in part 
on the Supercomputer Center, ISSP, University of Tokyo.

\clearpage
\appendix
\section{The pseudosphere}

The pseudosphere is defined as one sheet of the double-sheeted hyperboloid \cite{Coxeter}
\begin{equation}
x^2 + y^2 - z^2 = -1,
\label{eqa_01}
\end{equation}
possessing the Minkowskian metric $ds^2= dx^2 + dy^2 - dz^2$.
Since (\ref{eqa_01}) specifies the locus of points whose squared distance
from the origin are equal to $-1$,
it is called a {\it pseudo}sphere having the radius ${\rm i}=\sqrt{-1}$
by analogy with the sphere.

While the above definition is rigorous,
it is clumsy for computations since three coordinates are used for only two degrees
of freedom.
This cumbrousness is removed by using an alternative representation of the pseudosphere,
called the Poincar\'e disk model \cite{Balazs}.
Suppose that the upper hyperboloid sheet is projected onto the $x$-$y$ plane
using the following mapping:
\begin{equation}
(x,y,z) \to \left(\frac{x}{1+z}, \; \frac{y}{1+z} \right).
\end{equation}
This transforms the upper sheet to a unit circle endowed with the metric
\begin{equation}
ds^2 = w^2 (dx^2 + dy^2), \quad w=\frac{2}{1-x^2-y^2}.
\label{eqa_03}
\end{equation}
The unit circle possessing the metric (\ref{eqa_03})
is referred to as a Poincar\'e disk,
and it serves as a compact representation of the pseudosphere.
The boundary of the disk corresponds to the points at infinity of the hyperboloid.
The Gaussian curvature $\kappa$ on the disk is calculated
using the formula:
\begin{equation}
\kappa= -\frac{1}{w^2}
\left( \frac{\partial^2}{\partial x^2}+ \frac{\partial^2}{\partial y^2} \right) \ln w.
\label{eqa_04}
\end{equation}
From (\ref{eqa_03}) and (\ref{eqa_04}), 
we see that $\kappa=-1$ at arbitrary points on the disk.
Thus it follows that the pseudosphere is a curved surface 
with a constant negative curvature
$\kappa=-1$.


\end{document}